# Circuital model for the Maxwell Fish Eye perfect drain

Juan C. González, Dejan Grabovičkić, Juan C. Miñano, Pablo Benítez

Universidad Politécnica de Madrid, Cedint, Campus de Montegancedo, 28223 Madrid, Spain.

**Abstract.** Perfect drain for the Maxwell Fish Eye (MFE) is a non-magnetic dissipative region placed in the focal point to absorb all the incident radiation without reflection or scattering. The perfect drain was recently designed as a material with complex permittivity ε that depends on frequency. However, this material is only a theoretical material, so it can not be used in practical devices. Recently, the perfect drain has been claimed as necessary to achieve super-resolution [Leonhard 2009, *New J. Phys.* **11** 093040], which has increased the interest for practical perfect drains suitable for manufacturing. Here, we analyze the super-resolution properties of a device equivalent to the MFE, known as Spherical Geodesic Waveguide (SGW), loaded with the perfect drain. In the SGW the source and drain are implemented with coaxial probes. The perfect drain is realized using a circuit (made of a resistance and a capacitor) connected to the drain coaxial probes. Super-resolution analysis for this device is done in Comsol Multiphysics. The results of simulations predict the super-resolution up to $\lambda/3000$ and optimum power transmission from the source to the drain.

## 1. Introduction

"Super-resolution" stands for the capacity of an optical system to produce images with details below the classic Abbe diffraction limit. In the last decade super-resolution has been shown experimentally with devices made of left-handed materials [1][2] (that is, materials with negative dielectric and magnetic constants)[3][4]. Unfortunately, high absorption and small (wavelength scale) source-to-image distance are both present in these experiments. Nevertheless, these devices have been claimed to reach the theoretical limit of infinite resolution [3].

An alternative device for perfect imaging has recently been proposed [6][7]: the Maxwell Fish Eye (MFE) lens. Unlike previous perfect imaging devices, MFE uses materials with a positive, isotropic refractive index distribution. This device is very well known in the framework of Geometrical Optics because it is an Absolute Instrument [5], so every object point has a stigmatic image point.

Leonhardt [6] analyzed Helmholtz wave fields in the MFE lens in two dimensions (2D). These Helmholtz wave fields describe TE-polarized modes in a cylindrical MFE, i.e., modes in which electric field vector points orthogonally to the cross section of the cylinder. Leonhardt found a family of Helmholtz wave fields which have a monopole asymptotic behavior at an object point as well as at its stigmatic image point. Each one of these solutions describes a wave propagating from the object point to the image point. It coincides asymptotically with an outward (monopole) Helmholtz wave at the object point, as generated by a point source, and with an inward (monopole) wave at the image point, as it was sunk by an "infinitely-well localized drain" (which we call a "perfect point drain"). This perfect point drain absorbs the incident wave, with no reflection or scattering. This result has also been confirmed via a different approach [8].

The physical significance of a passive perfect point drain has been controversial [9]-[18]. In references [6] and [5] the perfect point drain was not physically described, but only considered as a mathematical entity (a point drain is represented by Dirac-delta as the point source).

However, a rigorous example of a passive perfect point drain for the MFE has recently been found, clarifying the controversy [19]. It consists of a dissipative region whose diameter tends towards zero and whose complex permittivity $\varepsilon$ takes a specific value depending on the operation frequency.

Two sets of experiments have recently been carried out to support the super-resolution capability in the MFE. In the first one, super-resolution with positive refraction has been demonstrated for the very first time at a microwave-frequency ($\lambda$=3 cm) [20][21]. The experimental results showed that two sources with a distance of $\lambda/5$ from each other (where $\lambda$ denotes the local wavelength $\lambda = \lambda_0/n$) could be resolved with an array made up of 10 drains spaced $\lambda/20$, which exceeded the ~$\lambda/2.5$ classic diffraction limit. Results with closer sources were not reported, but it should be noted that this experiment was limited to the resolution of the array of drains.

The second set of experiments has been carried out for the near infrared frequency ($\lambda = 1.55$ µm), but resolution below the diffraction limit was not found [22]. The authors assume that the failure in the experimental demonstration is due to manufacturing flaws in the prototype.

Although the perfect drain has not been used in these experiments, i.e. there was a reflected wave from the drain to the source, the MFE has shown super-resolution for microwave frequencies. This means that the perfect drain is not necessary for reaching the super-resolution (see also [23]). However, in this paper we will show that the use of the perfect drain increases super-resolution.

Recently, a device equivalent to the MFE, Spherical Geodesic Waveguide (SGW) made for microwave frequencies has been presented [23] [24]. The SGW is a spherical waveguide filled with a non-magnetic material and isotropic refractive index distribution proportional to $1/r$ ($\varepsilon = (r_0/r)^2$ and $\mu = 1$), $r$ being the distance to the center of the spheres. Transformation Optics theory [25] proves that the TE-polarized electric modes of the cylindrical MFE are transformed into radial-polarized modes in the SGW, so both have the same imaging properties. When the waveguide thickness is small enough, the variation of the refractive index within the two spherical shells can be ignored resulting in a constant refractive index within the waveguide. In [23] the SGW has been analyzed using two coaxial probes (source and drain) loaded with the characteristic impedances. The results have shown the super-resolution up to $\lambda/500$ for a discrete number of frequencies, called notch frequencies, that are close to the well known Schumman Resonance frequencies of spherical systems. For other frequencies the system did not present resolution below diffraction limit. In these analysis the perfect drain has not been used, thus beside the incident wave, a reflected wave existed in the SGW as well. However, the super-resolution properties have been shown.

Herein, we present an improvement of the super-resolution using the SGW with the perfect drain. The perfect drain is realized using a circuit (made of a resistance and a capacitor) connected to the drain coaxial probe. The difference between the presented drain and the perfect drain proposed in [19] is the practical realization. While in [19] the perfect drain is made of a material with complex permittivity $\varepsilon$, here it is only a coaxial line loaded with a resistor and a capacitor of conventional values (for example, $R$=2.57$\Omega$ and $C$= 55.05pF for $f$=0.25 GHz). Using the circuital model for the perfect drain, the Comsol simulations have shown the super-resolution up to $\lambda/3000$ for the same discrete frequencies as in [23], which is much higher than the $\lambda/500$ obtained without perfect drain.

In section 2, it is described the complete microwave circuits. In section 3 modal analysis of the SGW is made including: the rigorous procedure used to find the perfect drain, the analysis of the transmitted and evanescent modes and the concept of voltage and current wave in the SWG. Discussion and conclusions are presented in section 4.

## 2    Microwave circuit and parameters of the simulation.

The SGW is bounded by two metallic spherical shells. The media between the shells is air. Two coaxial probes have been used to simulate the source and drain in the SGW. Consider the microwave circuit consists of the generator $V_g$ with the impedance $Z_g$ (on the source port side), coaxial lines, the SGW, and the load with the impedance $Z_L$ (on the drain port side), as shown in Fig. 1.

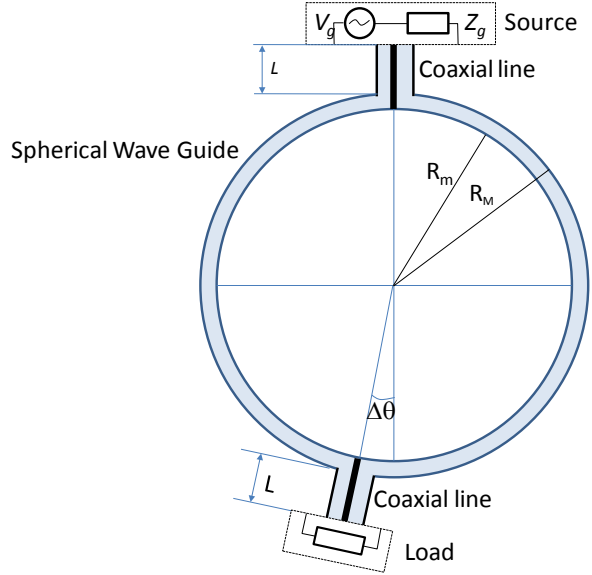

Fig. 1 Complete microwave circuit analyzed in this paper formed by: the source ($V_g$ and $Z_g$) connected to a coaxial transmission line of length $L$, the load connected ($Z_L$) to other identical transmission line and the spherical waveguide. $R_M$ and $R_m$ are radius of the external and internal metallic spheres.

The same circuit has been analyzed in [23] with the condition $Z_L=Z_g=Z_0$, where $Z_0$ is the characteristics impedance of the coaxial line, so we have:

$$|S_{21}|^2 = \frac{P_{load}}{P_{max}} \qquad (1)$$

Where, $S_{21}$ is the scattering parameter of the circuit [26], $P_{load}$ is the power delivered to the load $Z_L$ and $P_{max}$ is the maximum power that can be delivered by the generator. In [23] the results have been presented using function $|S_{21}|^2$ obtained for different frequencies and displacements $\Delta\theta$.

Here, in Section 4, it is repeated the same procedure using the impedance that simulates the perfect drain (it will be calculated in Section 3 and will be called $Z_{pd}$). The circuit is designed in Comsol with the conditions $Z_L=Z_g=Z_{pd}$ (see Fig. 1). The super-resolution is analyzed via function $P_{load}/P_{max}$ for different frequencies and displacements. Note, that now $P_{load}/P_{max}$ is different from $|S_{21}|^2$.

## 3  Modal analysis of the structure and numerical procedure to find the perfect drain.

The perfect drain, defined by the impedance $Z_L$, absorbs all the incident radiation without reflection inside the SWG when the source and drain are placed in opposite pole ($\Delta\theta=0$). In this section we present a rigorous procedure to find this impedance for a wide band of frequencies.

The structure does not depend on the cylindrical (coaxial) and spherical (SWG) coordinate $\phi$, thus:

- The unique modes of the coaxial guide without $\phi$ dependence are the TEM modes, so in the interface between the coaxial and sphere only exist the TEM modes, incident and reflected [26].

- Inside the SGW the electric and magnetic fields are necessary of the form:

$$\mathbf{E}(r,\theta) = E_r(r,\theta)\mathbf{r} + E_\theta(r,\theta)\boldsymbol{\theta}$$
$$\mathbf{H}(r,\theta) = H_\phi(r,\theta)\boldsymbol{\phi} \qquad (2)$$

The complete analysis of the modes is done using the same procedure explained by Wu in [27] on a radial-line/coaxial-line junction. Fig. 2 shows the complete region of the junction separated into three regions: the coaxial (region 1), the SGW (region 2) and the common region (region 3). The electronic field is calculated in each of these three regions using the procedure described below.

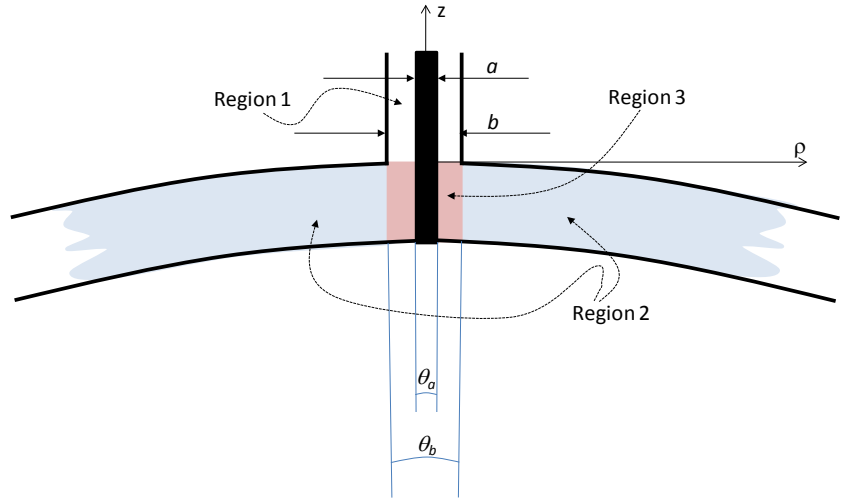

Fig. 2 Three regions of the junction used for the analysis. The coaxial is the white region, the SGW the blue and the common region is the read. $\theta_a$ and $\theta_b$ are the angles in spherical coordinates covered by the inner and outer conductor of the coaxial.

The dimension of the structure analyzed here are (Fig. 1 and Fig. 2):

- $R_M$=1005 mm.
- $R_m$=1000 mm.
- $a$=5 mm ($\theta_a$= 0.285°)
- $b$=10 mm ($\theta_b$=0.57°)
- $L$=20 mm.

### 3.1. Field in the coaxial line (region 1).

Due to the symmetry of the structure, the field of the coaxial probe is completely described by the TEM modes of the line [26]:

$$\mathbf{E} = \left(\frac{V_e^+}{\rho Ln(b/a)} e^{-jk_o z} + \frac{V_e^-}{\rho Ln(b/a)} e^{jk_o z}\right)\boldsymbol{\rho}$$

$$\mathbf{H} = \frac{1}{\sqrt{\mu/\varepsilon}} \left(\frac{V_e^+}{\rho Ln(b/a)} e^{-jk_o z} - \frac{V_e^-}{\rho Ln(b/a)} e^{jk_o z}\right)\boldsymbol{\phi} \qquad (3)$$

Where *a* and *b* are the diameters of the internal and external coaxial (Fig. 2), $\rho$ and *z* the cylindrical coordinates, $k_o$ the propagation constant and $V_e^+$ and $V_e^-$ constants. With the condition $b \ll R_M$, on the surface $z=0$ (Fig. 1 and Fig. 2), the fields can be approximated to:

$$\mathbf{E} = \left(\frac{V_e^+}{R_M \sin(\theta) Ln(b/a)} + \frac{V_e^-}{R_M \sin(\theta) Ln(b/a)}\right)\boldsymbol{\theta}$$

$$\mathbf{H} = \frac{1}{\sqrt{\mu/\varepsilon}}\left(\frac{V_e^+}{R_M \sin(\theta) Ln(b/a)} - \frac{V_e^-}{R_M \sin(\theta) Ln(b/a)}\right)\boldsymbol{\varphi}$$

(4)

Where $\theta$ and $\varphi$ are spherical coordinates.

### 3.2. Field in the SGW (region 2).

With the condition $R_M - R_m \ll R_M$ one complete set of solutions inside the SWG fulfilling the boundary condition of metallic surfaces are [28]:

$$E_r(r,\theta) = \sum_n \left[ A_n F_{v_n}(\cos(\theta)) + B_n R_{v_n}(\cos(\theta)) \right] \left( \left(\frac{n\pi}{R_M - R_m}\right)^2 + k_0^2 \right)\left(-\cos\left(\frac{(r-R_m)n\pi}{R_M - R_m}\right)\right)$$

$$E_\theta(r,\theta) = \sum_n \frac{1}{r}\left[ A_n F_{v_n}'(\cos(\theta)) + B_n R_{v_n}'(\cos(\theta)) \right]\frac{-n\pi}{R_M - R_m}\sin\left(\frac{(r-R_m)n\pi}{R_M - R_m}\right)$$

$$H_\varphi(r,\theta) = -j\omega\varepsilon_0\sum_n \left[ A_n F_{v_n}'(\cos(\theta)) + B_n R_{v_n}'(\cos(\theta)) \right]\frac{1}{r}\cos\left(\frac{(r-R_m)n\pi}{R_M - R_m}\right)$$

$$v_n \approx -0.5 + 0.5\sqrt{1 + 4\left((R_M k_o)^2 - \left(\frac{R_M n\pi}{R_M - R_m}\right)^2\right)}$$

(5)

Where $A_n$ and $B_n$ are constants and $F_{vn}$ and $R_{vn}$, are called Forwards and Reverse functions defined as:

$$F_{v_n}(x) = P_{v_n}(x) + j\frac{2}{\pi}Q_{v_n}(x)$$

$$R_{v_n}(x) = P_{v_n}(x) - j\frac{2}{\pi}Q_{v_n}(x)$$

(6)

$P_{vn}$ and $Q_{vn}$, are the Legendre function of first and second kind.

### 3.3. Field in the common region of coaxial and SGW (region 3).

The solution in this region has the same form as the solution expressed in Eq.(5), but now having an additional term, one particular solution, necessary to fulfill the boundary conditions at the common surfaces.

$$E_r(r,\theta) = \sum_n \left[ D_n F_{v_n}(\cos(\theta)) + E_n R_{v_n}(\cos(\theta)) \right] \left( \left(\frac{n\pi}{R_M - R_m}\right)^2 + k_0^2 \right) \left( -\cos\left(\frac{(r-R_m)n\pi}{R_M - R_m}\right) \right)$$

$$E_\theta(r,\theta) = E_p(r,\theta) + \sum_n \frac{1}{r} \left[ D_n F'_{v_n}(\cos(\theta)) + E_n R'_{v_n}(\cos(\theta)) \right] \frac{-n\pi}{R_M - R_m} \sin\left(\frac{(r-R_m)n\pi}{R_M - R_m}\right) \quad (7)$$

$$H_\varphi(r,\theta) = H_p(r,\theta) - j\omega\varepsilon_0 \sum_n \left[ D_n F'_{v_n}(\cos(\theta)) + E_n R'_{v_n}(\cos(\theta)) \right] \frac{1}{r} \cos\left(\frac{(r-R_m)n\pi}{R_M - R_m}\right)$$

Where the particular solution $E_p$ and $H_p$ has to fulfill the Maxwell equations and the boundary conditions at $r=R_M$ and $r=R_m$.

$$\mathbf{E_p} = E_p(r,\theta)\mathbf{\theta} \qquad \nabla \times \mathbf{E_p} = \frac{1}{r}\frac{\partial}{\partial r}(rE_p(r,\theta))\mathbf{\varphi} = -j\omega\mu\mathbf{H_p}(r,\theta)$$

$$\nabla \times \mathbf{H_p} = \frac{-1}{j\omega\mu}\frac{1}{r^2 \sin(\theta)}\frac{\partial}{\partial\theta}\left(\sin(\theta)\frac{\partial}{\partial r}(rE_p(r,\theta))\right)\mathbf{r} + \qquad (8)$$

$$\frac{1}{j\omega\mu}\frac{1}{r}\frac{\partial}{\partial r}\left(\frac{\partial}{\partial r}(rE_p(r,\theta))\right)\mathbf{\theta} = j\omega\varepsilon E_p(r,\theta)\mathbf{\theta}$$

From the second equation in Eq. (8), it is necessary:

$$E_p(r,\theta) = f(r)\frac{1}{\sin(\theta)}$$

$$\frac{d^2}{dr^2}(rf(r)) = -k_0^2 rf(r) \qquad (9)$$

$$k_0^2 = \omega^2\mu\varepsilon$$

Solving the differential equation from Eq. (9), the particular solution for the fields $E_p$ an $H_p$ are obtained:

$$E_p(r,\theta) = \frac{a_1 \frac{e^{jk_0 r}}{r} + a_2 \frac{e^{-jk_0 r}}{r}}{\sin(\theta)} \qquad H_p(r,\theta) = \frac{-jk_0}{j\omega\mu} \frac{a_1 \frac{e^{jk_0 r}}{r} - a_2 \frac{e^{jk_0 r}}{r}}{\sin(\theta)} \qquad (10)$$

Where $a_1$ and $a_2$ are two integration constants. These constants are obtained using the boundary conditions for $E_p$ at $r=R_M$ (Eq. (4)) and $r=R_m$ ($E_p=0$ for metallic surface):

$$a_1 = (V_e^+ + V_e^-)\frac{1}{Ln(b/a)}\left(\frac{1}{e^{jk_0 R_M} - e^{jk_0(2R_m - R_M)}}\right)$$

$$a_2 = (V_e^+ + V_e^-)\frac{1}{Ln(b/a)}\left(\frac{-e^{jk_0 2R_m}}{e^{jk_0 R_M} - e^{jk_0(2R_m - R_M)}}\right) \qquad (11)$$

The fields defined in Eq. (10) and Eq. (11) can be calculated using a series expansion as follow:

$$E_p(r,\theta) = \frac{V_e^+ + V_e^-}{r\sin(\theta)} \sum_n d_n \frac{-n\pi}{R_M - R_m} \sin[\frac{(r-R_m)n\pi}{R_M - R_m}]$$

$$H_p(r,\theta) = \frac{V_e^+ + V_e^-}{r\sin(\theta)} \sum_n e_n \cos[\frac{(r-R_m)n\pi}{R_M - R_m}]$$

(12)

Where $d_n$ and $e_n$ are the expansion constants satisfying $j\omega\varepsilon d_n = -e_n$. The complete field in this region is then defined as follow:

$$E_r(r,\theta) = \sum_n \left[ D_n F_{v_n}(\cos(\theta)) + E_n R_{v_n}(\cos(\theta)) \right] \left( \left(\frac{n\pi}{R_M - R_m}\right)^2 + k_0^2 \right) (-\cos(\frac{(r-R_m)n\pi}{R_M - R_m}))$$

$$E_\theta(r,\theta) = \sum_n \frac{1}{r} \left[ D_n F'_{v_n}(\cos(\theta)) + E_n R'_{v_n}(\cos(\theta)) + \frac{V_e^+ + V_e^-}{\sin(\theta)} \frac{e_n}{-j\omega\varepsilon_0} \right] \frac{-n\pi}{R_M - R_m} \sin(\frac{(r-R_m)n\pi}{R_M - R_m})$$  (13)

$$H_\varphi(r,\theta) = -j\omega\varepsilon_0 \sum_n \left[ D_n F'_{v_n}(\cos(\theta)) + E_n R'_{v_n}(\cos(\theta)) + \frac{V_e^+ + V_e^-}{\sin(\theta)} \frac{e_n}{-j\omega\varepsilon_0} \right] \frac{1}{r} \cos(\frac{(r-R_m)n\pi}{R_M - R_m})$$

In the coaxial line (on the drain side) it can be made the same development with the results:

$$E_r(r,\theta) = \sum_n \left[ G_n F_{v_n}(\cos(\pi-\theta)) + I_n R_{v_n}(\cos(\pi-\theta)) \right] \left( \left(\frac{n\pi}{R_M - R_m}\right)^2 + k_0^2 \right) (-\cos(\frac{(r-R_m)n\pi}{R_M - R_m}))$$

$$E_\theta(r,\theta) = \sum_n \frac{1}{r} \left[ G_n F'_{v_n}(\cos(\pi-\theta)) + I_n R'_{v_n}(\cos(\pi-\theta)) + \frac{V_s^+ + V_s^-}{\sin(\theta)} \frac{e_n}{-j\omega\varepsilon_0} \right] \frac{-n\pi}{R_M - R_m} \sin(\frac{(r-R_m)n\pi}{R_M - R_m})$$  (14)

$$H_\varphi(r,\theta) = -j\omega\varepsilon_0 \sum_n \left[ G_n F'_{v_n}(\cos(\pi-\theta)) + I_n R'_{v_n}(\cos(\pi-\theta)) + \frac{V_s^+ + V_s^-}{\sin(\theta)} \frac{e_n}{-j\omega\varepsilon_0} \right] \frac{1}{r} \cos(\frac{(r-R_m)n\pi}{R_M - R_m})$$

### 3.4. Transmitted and not transmitted modes.

Different modes inside the sphere are defined by $v_n = \alpha_n + j\beta_n$ (see Eq. (5)), which is in general a complex number. According to the asymptotic expression of Legendre functions $P_{vn}(x)$ and $Q_{vn}(x)$ for $x$ close to 1 and -1 the following results are obtained [29]:

$$\left|\frac{F_{v_n}(\cos(\pi-\theta))}{F_{v_n}(\cos(\theta))}\right| \to \frac{e^{-\beta_n}}{\Gamma(0)} \qquad \left|\frac{R_{v_n}(\cos(\theta))}{R_{v_n}(\cos(\pi-\theta))}\right| \to \Gamma(0)e^{-\beta_n} \quad for \quad \theta \to 0$$

$$\beta_n = \text{Im}[v_n] \qquad \Gamma(\theta), \text{ Gamma function}$$

(15)

Having in mind the SGW dimension and the frequencies of interest (the microwave frequencies from 0.2 GHz to 0.4 GHz) the parameter $v_n$ is real only for $n=0$. For example for f=300 MHz, $v_0$ =4.98, $v_1$ =-0.5+631.4j, $v_2$ =-0.5+1262.9j etc. Fig. 3 shows for example the graphs for Log($|F_{vn}(\cos(\theta))/F_{vn}(\cos(\theta_b))|$ ) and Log($|R_{vn}(\cos(\theta))/R_{vn}(\cos(\theta_b))|$ ) respect to $\theta$ for one complex $v_n$=-0.5+4.0j ($\theta_b$ is defined in Fig. 2). Clearly $|F_{vn}(\cos(\theta))|$ suffers a near exponential attenuation between $\theta=\theta_b$ and $\theta=\pi-\theta_b$ as shown in Eq.(15). The same occurs for $|R_{vn}(\cos(\theta))|$ between $\theta=\pi-\theta_b$ and $\theta=\theta_b$. The similar results are found for every complex $v_n$, so $F_{vn}(\cos(\theta))$ and $R_{vn}(\cos(\theta))$ can be considered evanescent waves. The SWG works as a single-mode with the electric and magnetic fields far for the interface region given by:

$$\mathbf{E}(r,\theta) = k_0^2 \left[ A_0 F_{v_0}(\cos(\theta)) + B_0 R_{v_0}(\cos(\theta)) \right] \mathbf{r}$$

$$\mathbf{H}(r,\theta) = -j\omega\varepsilon_0 \left[ A_0 F'_{v_0}(\cos(\theta)) + B_0 R'_{v_0}(\cos(\theta)) \right] \frac{1}{r} \boldsymbol{\phi} \quad (16)$$

The Poynting vector and transmitted power through a surface, defined by $\theta$=cte, in the direction $\theta$ have the following form:

$$\mathbf{S}(\theta) = \frac{1}{2}(\mathbf{E}\times\mathbf{H}^*) =$$

$$k_0^2 j\omega\varepsilon (A_0 F_v(\cos(\theta)) + B_0 R_v(\cos(\theta)))(A_0 \frac{dF_v(\cos(\theta))}{d\theta} + B_0 \frac{dR_v(\cos(\theta))}{d\theta})^* \boldsymbol{\theta} \quad (17)$$

$$P = \mathrm{Re}[\iint_{\substack{R_m < r < R_M \\ 0 < \varphi < 2\pi}} S(\theta) r \sin(\theta) dr d\varphi] = 2k_0^2 \omega\varepsilon (R_M - R_m)(|A_0|^2 - |B_0|^2)$$

This equation explains clearly the concept of $F_{v0}$ and $R_{v0}$ as incident and reflected waves.

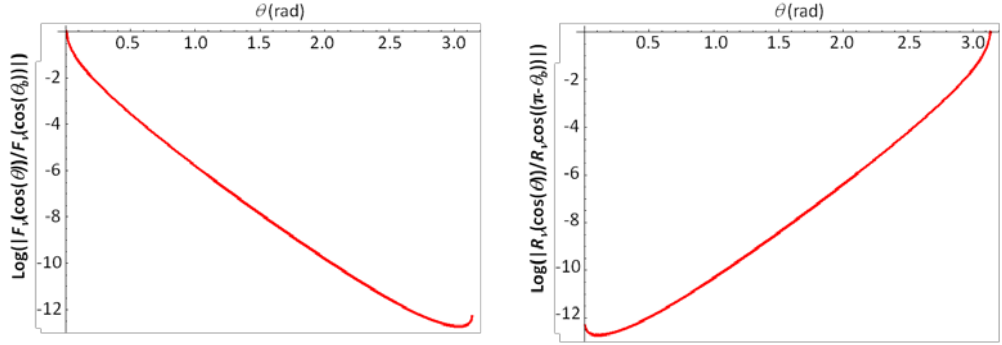

Fig. 3 Graphs for Log($|F_{vn}(\cos(\theta))/F_{vn}(\cos(\theta_b))|$) and Log($|R_{vn}(\cos(\theta))/R_{vn}(\cos(\pi-\theta_b))|$) as function of $\theta$ for $v_n$=-0.5+4.0j. $F_{vn}(\cos(\theta))$ and $R_{vn}(\cos(\theta))$ have a near exponential attenuation similar to the evanescent waves in conventional waveguides. The modes with complex $v_n$ are not transmitted. The only modes that exist inside the SWG far from the interface are the modes having real $v_n$. For the band of frequencies analyzed here only $v_0$ is real and the guide works as single-mode.

### 3.5. Boundary conditions.

The coefficients $A_n$ and $B_n$ from Eq. (5), $D_n$ and $E_n$ from Eq. (7) and $G_n$ and $I_n$ from Eq. (14) are obtained using the following boundary conditions (Fig. 2):
- Tangent electric field in the inner conductor is null.
- Electric and magnetic field are the same in the common surface between regions 1 and 2.
- The same conditions in the coaxial at the drain side.

$$D_n F_{v_n}(\cos(\theta_a)) + E_n R_{v_n}(\cos(\theta_a)) = 0$$

$$D_n F_{v_n}(\cos(\theta_b)) + E_n R_{v_n}(\cos(\theta_b)) = A_n F_{v_n}(\cos(\theta_b)) + B_n R_{v_n}(\cos(\theta_b))$$

$$D_n F'_{v_n}(\cos(\theta_b)) + E_n R'_{v_n}(\cos(\theta_b)) + \frac{V_e^+ + V_e^-}{\sin(\theta_b)} \frac{e_n}{-j\omega\varepsilon_0} = A_n F'_{v_n}(\cos(\theta_b)) + B_n R'_{v_n}(\cos(\theta_b)) \quad (18)$$

$$G_n F_{v_n}(\cos(\pi-\theta_a)) + I_n R_{v_n}(\cos(\pi-\theta_a)) = 0$$

$$G_n F_{v_n}(\cos(\pi-\theta_b)) + I_n R_{v_n}(\cos(\pi-\theta_b)) = A_n F_{v_n}(\cos(\pi-\theta_b)) + B_n R_{v_n}(\cos(\pi-\theta_b))$$

$$G_n F'_{v_n}(\cos(\pi-\theta_b)) + I_n R'_{v_n}(\cos(\pi-\theta_b)) + \frac{V_s^+ + V_s^-}{\sin(\pi-\theta_b)} \frac{e_n}{-j\omega\varepsilon_0} = A_n F'_{v_n}(\cos(\pi-\theta_b)) + A_n R'_{v_n}(\cos(\pi-\theta_b))$$

Where $\theta_a$ and $\theta_b$ are defined in Fig. 2. When the voltage in the two coaxial lines ($V_e=V_e^+ +V_e^-$ and $V_s=V_s^+ +V_s^-$ ) are known, the system (18) can be solved for each $n$.

### 3.6. Perfect drain.

According the previous analysis the SGW work as a single-mode guide, so the condition for the perfect drain (there is no reflected wave in the guide) is satisfied when $B_0=0$ in Eq. (16). The procedure to obtain the perfect drain consists of the following steps:
- In Eq.(18) for $n=0$ it is imposed the condition $B_0=0$. Then $A_0$, $D_0$, $E_0$, $G_0$, $I_0$ and $V_s^+ +V_s^-$ are calculated. $V_s^+ +V_s^-$ is the necessary voltage (on the drain side coaxial) for perfect wave absorption in the SWG.
- The coefficient $A_n$, $B_n$, $D_n$, $E_n$, $G_n$ and $I_n$ are obtained using Eq. (18) and the voltage $V_s^+ +V_s^-$ calculated in the previous step.
- The field $H_\varphi$ for $r=R_M$ and $\theta_a < \theta < \theta_b$ is computed using Eq. (13).
- For the sake of uniqueness of solution, this field has to depend on $\theta$ as in Eq.(4). The voltage $V_s^+ -V_s^-$ is then obtained.
- The impedance of the coaxial line at $r=R_M$ and the load are defined as:

$$Z_S = Z_0 \frac{V_s^+ +V_s^-}{V_s^+ -V_s^-} \qquad Z_L = Z_0 \frac{V_s^+ e^{jk_oL} +V_s^- e^{-jk_oL}}{V_s^+ e^{jk_oL} -V_s^- e^{-jk_oL}} \tag{19}$$

Fig. 4 shows the real and imaginary parts of the perfect drain impedance calculated for a band of frequencies of interest, the results show some oscillation due numerical error in the calculation. Linear approximation is presented. Fig. 5 shows the comparison between the module of the theoretical electric field (in the case of the perfect drain, there exit only forward wave, given by the function $F_{v0}(\theta)$) and the module of the electric field simulated in COMSOL. Although the theoretical value of the electric field at $\theta=0$ and $\theta=\pi$ is infinite (due to the properties of the Legendre functions), for the sake of clearness of the graphs in Fig. 5, we did not extend the theoretical graph to these points. Fig. 5 shows perfect matching between the theoretical and simulated fields for $f=0.25$ GHz. The circuit parameters are calculated using the linear approximations presented in Fig. 4, $R=2.57\Omega$ and $C=55.05$pF.

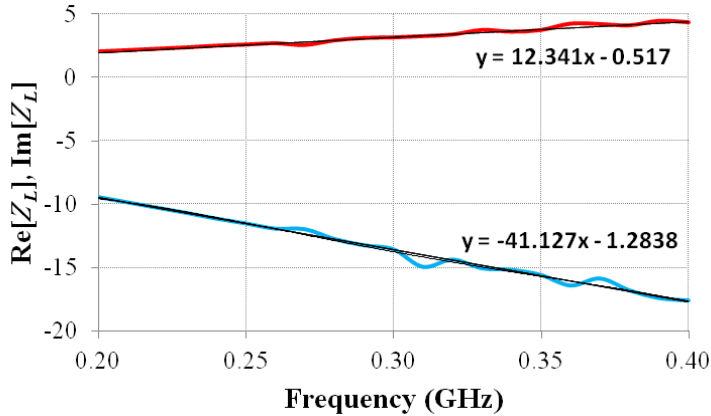

Fig. 4 Real (in red) and imaginary (in blue) parts of the perfect drain impedance for different frequencies.

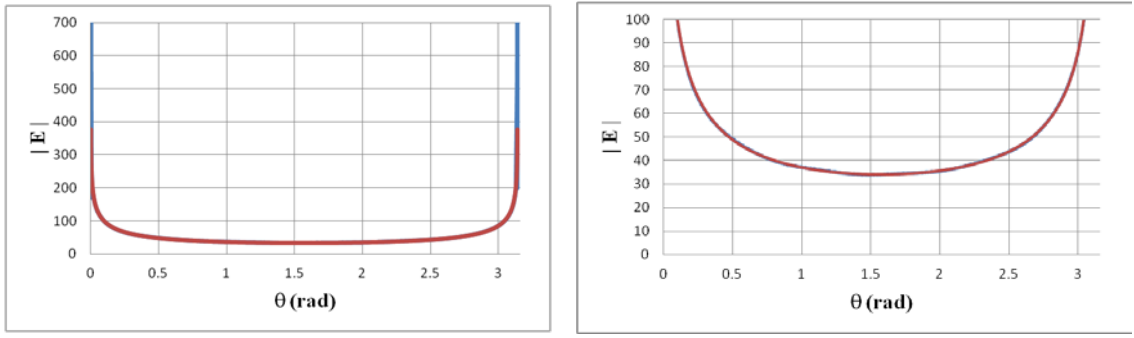

Fig. 5 Left, The module of the theoretic electric field (in red) and the modul of the electric filed simulated in Comsol (in blue) for the SGW with the perfect drain. Right, close up of the same graphs.

### 3.7. Concept of voltage and current wave in the spherical waveguide

When the spherical guide works as a single-mode guide, then the system can be analyzed using the concept of voltage and current wave, in similar manner as it is used in conventional guides. From Eq. (5) when $n=0$, the fields are:

$$E_r(r,\theta) = k_0^2 \left[ D_0 F_{v_0}(\cos(\theta)) + E_0 R_{v_0}(\cos(\theta)) \right]$$
$$H_\varphi(r,\theta) = -j\omega\varepsilon_0 \frac{1}{r} \left[ D_0 \frac{dF_{v_0}(\cos(\theta))}{d\theta} + E_0 \frac{dR_{v_0}(\cos(\theta))}{d\theta} \right] \quad (20)$$

Incident and reflected voltage and current wave are defined as follow:

$$V_i(\theta) = \int_{R_m}^{R_M} k_0^2 A F_v(\cos(\theta)) dr = k_0^2 A (R_M - R_m) F_v(\cos(\theta))$$
$$V_r(\theta) = \int_{R_m}^{R_M} k_0^2 A R_v(\cos(\theta)) dr = k_0^2 B (R_M - R_m) R_v(\cos(\theta)) \quad (21)$$
$$I_i(\theta) = \int_0^{2\pi} \frac{-j\omega\varepsilon_0}{r} A \frac{dF_v(\cos(\theta))}{d\theta} r \sin(\theta) d\varphi = -j\omega\varepsilon_0 2\pi A \frac{dF_v(\cos(\theta))}{d\theta} \sin(\theta)$$
$$I_r(\theta) = \int_0^{2\pi} \frac{-j\omega\varepsilon_0}{r} A \frac{dR_v(\cos(\theta))}{d\theta} r \sin(\theta) d\varphi = -j\omega\varepsilon_0 2\pi B \frac{dR_v(\cos(\theta))}{d\theta} \sin(\theta)$$

The transmitted power expressed in (17) can be obtained from (21) as:

$$P = \frac{1}{2} \text{Re}[VI^*] = \frac{1}{2} \text{Re}[(V_i(\theta) + V_r(\theta))(I_i(\theta) + I_r(\theta))^*] \quad (22)$$

## 4 SR analysis of the SGW matched with the perfect drain.

The SGW with the perfect drain is designed and analyzed in Comsol. In order to show super-resolution properties of the SGW, we have made several simulations for different displacements of the drain port, and for different values of frequency. Special care has been taken to define the mesh of the system. In order to mesh the guide properly, the geometry has been divided into few domains. Each domain is meshed separately according to its geometric and physical properties. Since the guide thickness is very low $(R_M-R_m)/R_m \ll 1$, the SGW is meshed using a swept mesh

(2D triangular mesh from the outer surface is swept to the inner surface, see Fig. 6 b)). On the other side, the coaxial cables are meshed with higher density using 3D tetrahedra. The mesh density is increased since the change of the electric field is significant in the neighbourhood of the coaxial cables.

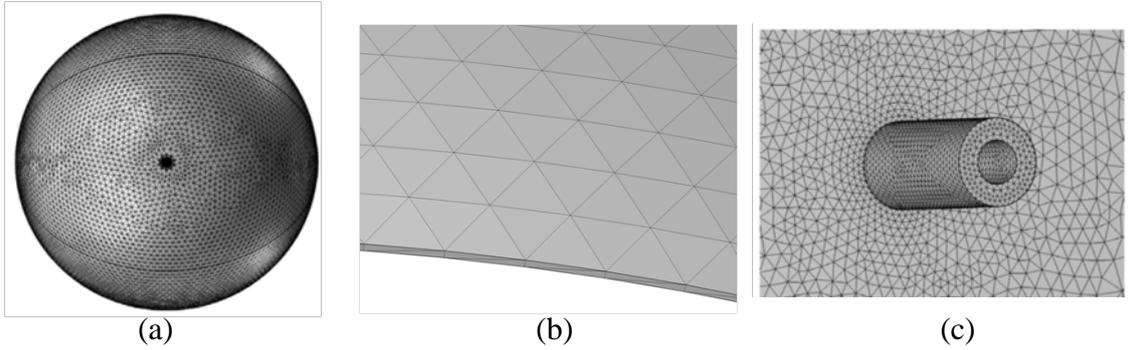

(a)          (b)          (c)

Fig. 6 Mesh structure. (a) SGW with coaxial cables, (c) close up of a piece of spherical shells, (c) close up of one coaxial cable from outside the sphere

### 4.1. $P_{load}/P_{max}$ as function of frequencies for different source port positions.

We have computed $P_{load}/P_{max}$ using COMSOL for a frequency range between 0.2 GHz and 0.4 GHz for different positions of the drain port. The source port is fixed at the source's image point, that is, $\theta = 0$, while the drain port is shifted $\lambda/N$ (for $\lambda=1$m corresponding to 0.3 GHz and N>100) in the neighborhood of the image point, that is, $\theta = \pi$ (see Fig. 1 ). When the drain port is placed in the image point, all the power is delivered perfectly. This can be achieved for all the frequencies using corresponding perfect drain impedance (Fig. 4). However, when the drain is moved from the image point, some of the power reflects, so the power delivered to the drain decreases. This power drop is extremely abrupt for some frequencies very close to the Schumann frequencies, called notch frequencies (see also [23]). Fig. 7 shows $P_{load}/P_{max}$ as function of the frequency in a narrow band around a notch frequency for different drain port positions. The notches get wider when the drain port is shifted further from the image point of the source, but the null of $P_{load}/P_{max}$ remains fixed, (see Fig. 7). The frequencies corresponding to these nulls are called notch frequencies.

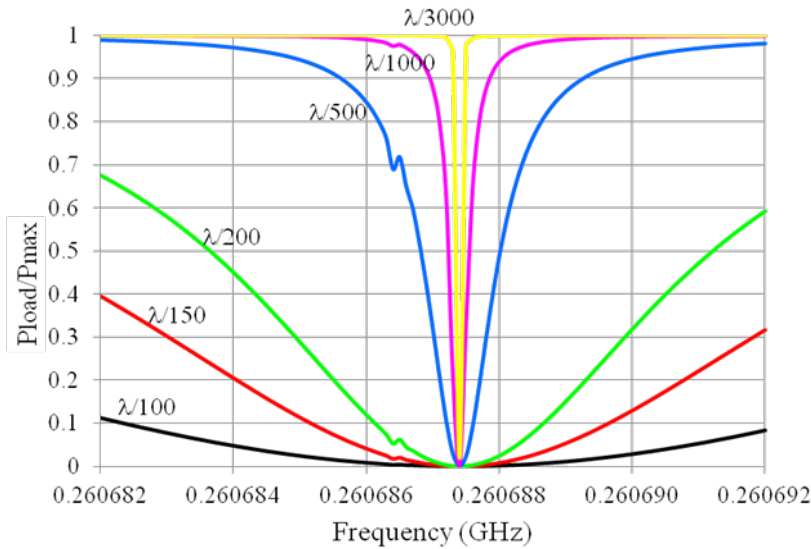

Fig. 7 Detail of $P_{load}/P_{max}$ as function of the frequency in a narrow band around a notch frequency for different drain port positions. The notch frequency is f=0.2606874 GHz

($\nu$=4.996). The nearest Schumann frequency is f=0.26086609 GHz ($\nu$=5) which is out the range of this Figure.

Fig. 7 shows $P_{load}/P_{max}$ for different drain port positions in a very narrow band in the neighborhood of the notch frequency close to $\nu_0$=5. The curves correspond to different shifts of the drain port. The shifts are in all cases much smaller than wavelength (from $\lambda$/100 to $\lambda$/3000 with $\lambda$=1.15084047 m that correspond to f=0.2606874 GHz, see Fig. 7). These results are quite surprising, since close to a specific frequency the power transmitted to the drain port suddenly reduces to a value near zero.

### 4.2. $P_{load}/P_{max}$ as function of drain port shift for different frequencies.

Since $P_{load}/P_{max}$ is proportional to the transmitted power, the graph representing $P_{load}/P_{max}$ versus the drain port shift (Fig. 8) is equivalent to the Point-Spread-Function (PSF) commonly used in Optics. This equivalence may seem surprising since the PSF is defined as the square of the electric field amplitude calculated in the absence of absorbers in the image space, and $P_{load}/P_{max}$ is defined in terms of the power transmitted to an absorber. However, the equivalence comes from the fact that, in Optics, the detection at the image is assumed to be made with a sensor which does not perturb the free-space fields; or that even if it does perturb the fields, it is assumed that still the sensor signal is proportional to the field amplitude (or its square, which is the PSF. Fig. 8 shows $P_{load}/P_{max}$ versus the drain port shifts for two frequencies. The blue curve corresponds to f=0.2847 GHz, i.e., far from a notch frequency ($\nu_0$ = 5.5).

Let us define "resolution" as the arc length (in wavelength units) that a drain port needs to be shifted so $P_{load}/P_{max}$ drops to 10% (not far from the Rayleigh criteria in Optics, which refers to the first null). With this definition, the diffraction limited resolution given by the blue curve is $\lambda$/3. The red curve corresponds to notch frequency f=0.26068741 GHz ($\nu_0$ =4.996) which clearly shows a much better resolution.

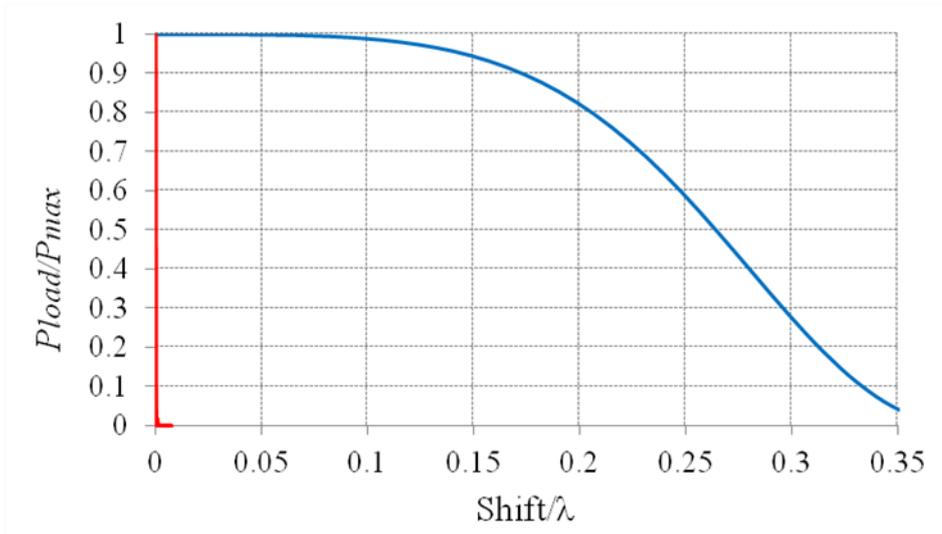

Fig. 8 $P_{load}/P_{max}$ as function of the drain port shift for a frequency near a notch one (red curve) and for a frequency far from the notch one (blue curve).

Fig. 9 is a blow-up of Fig. 8 in the upper neighborhood of a notch frequency. The graph for frequencies slightly below the notch frequency is similar. Note that Fig. 9 shows the same information as Fig. 7 but plotting $P_{load}/P_{max}$ vs. the drain port shift (expressed in units of $\lambda$) and using the frequency as a parameter.

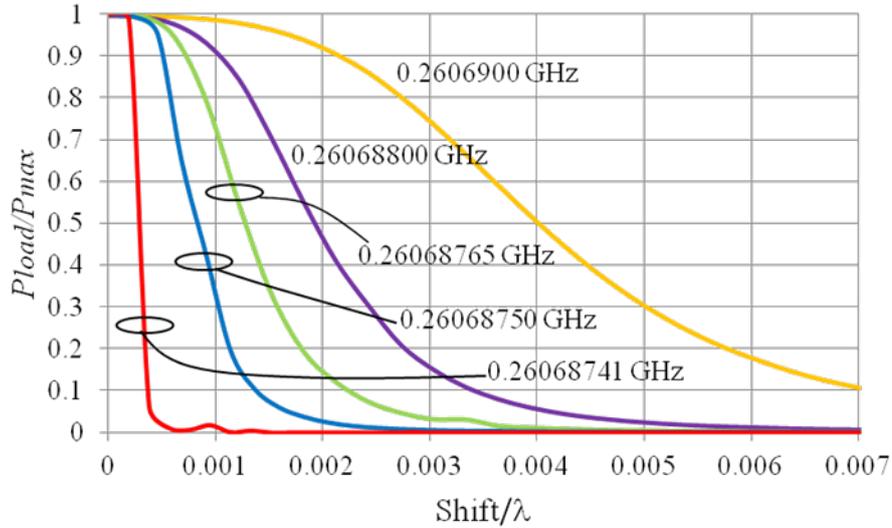

Fig. 9 $P_{load}/P_{max}$ as function of the drain port shift for different frequencies corresponding to super-resolution between $\lambda/3000$ and $\lambda/140$.

From the orange to the red curves, increasing resolutions are achieved: 0.007 $\lambda$ (that is, $\lambda/140$) for the orange to $\lambda/3000$ for the red. The latter, whose frequency $f$=0.26068741 GHz corresponds to $\nu$=4.99636) is the highest resolution that we have obtained. Computations for frequencies near the notch frequency show essentially null $P_{load}/P_{max}$ values for shifts > $\lambda/3000$ (as in the red line in the picture). $P_{load}/P_{max}$ values for shifts below $\lambda/3000$ (excepting no shift or shifts very near to zero) and frequencies near a notch frequency are inconsistent (the solver did not converge to a single solution due to numerical errors). It seems that Leonhardt's assertion of infinite resolution (i.e., perfect imaging) may occur for the discrete notch frequencies in the SGW, although the mentioned inconsistencies have prevented us from numerically predicting resolutions beyond $\lambda/3000$.

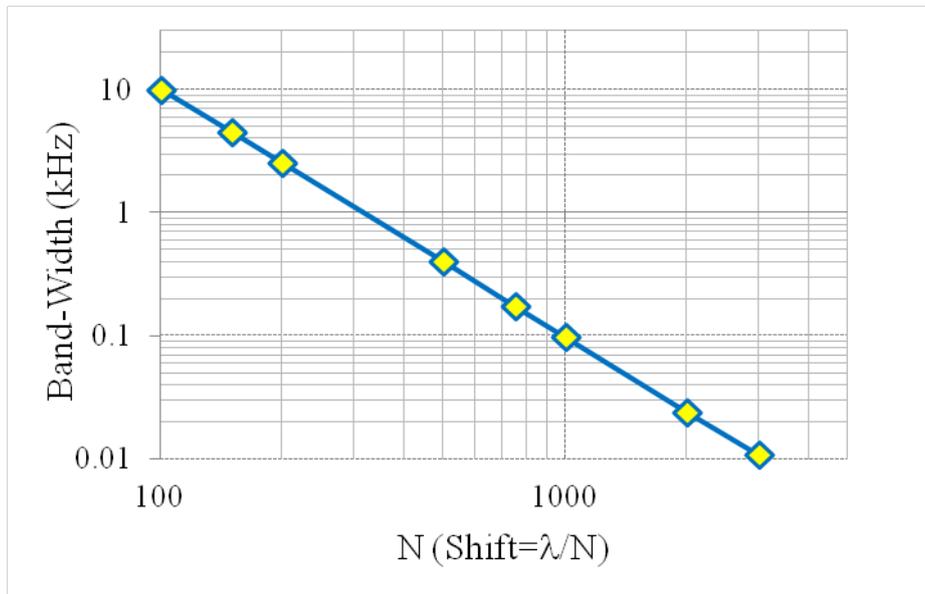

Fig. 10 Bandwidth as a function of the resolution. The abscissa axis shows $N$, meaning that the resolution is better than $\lambda/N$.

The $\lambda/3000$ resolution is achieved only for a narrow bandwidth ($\approx$ 10 Hz, which is much smaller than the notch frequency $\approx$0.3GHz). If larger bandwidths are needed, lower resolutions (but still sub-wavelength) may be achieved.

Fig. 10 shows the bandwidth vs. *N*, meaning that the resolution is better that $\lambda/N$. The bandwidth has been calculated as $f_{max} - f_{min}$ with $f_{max}$ and $f_{min}$ fulfilling $P_{load}/P_{max}(f_{max}) = P_{load}/P_{max}(f_{min}) = 0.1$, using the information of the curves in Fig. 7 and similar curves. The linear dependence shown in Fig. 10 (slope -2) reveals that the product $N^2 \times$ bandwidth is constant in the range analyzed here.

## 5 Discussion.

Leonhardt in [6] and [7] suggested that MFE should produce perfect imaging for any frequency using perfect drains. However, the experiments in [20][21] and simulations from [23], have shown super-resolution properties of the MFE, although the perfect drain has not been used. In these references, the coaxial probes were loaded with their characteristic impedances, so the absorption of the incident wave was not perfect. Leonhardt assumed that the ability of the MFE to propagate the wave, generated by a point source, toward to a perfect point drain was enough to guarantee perfect imaging. This does not seem to be sufficient, since it does not provide information on how much power the drain will absorb when it is displaced out of the image point. The simulations presented here show that super-resolution only happens for a particular set of frequencies known as notch frequencies, the same one as in [23]. The presented results have shown maximum super-resolution λ/3000, which is much higher than in the case where there were not perfect drains (λ/500, see [23]). Also, the frequency bandwidth has increased 20 times, e.g. for λ/500 the bandwidth is about 400 Hz (Fig. 10), while in [23] it was only 20 Hz.

## 6 References.